\documentclass[aps,prl,twocolumn,showpacs,preprintnumbers,amsmath,amssymb,superscriptaddress]{revtex4}%
\usepackage{graphicx}% Include figure files
\usepackage{dcolumn}% Align table columns on decimal point
\usepackage{bm}% bold math
\usepackage{color}

\begin{document}

\preprint{preprint(\today)}

\title{Magnetic field enhanced structural instability in EuTiO$_{3}$}

\author{Z.~Guguchia}
%\email{zurabgug@physik.uzh.ch} 
\affiliation{Physik-Institut der
Universit\"{a}t Z\"{u}rich, Winterthurerstrasse 190, CH-8057
Z\"{u}rich, Switzerland}

\author{H.~Keller}
\affiliation{Physik-Institut der Universit\"{a}t Z\"{u}rich,
Winterthurerstrasse 190, CH-8057 Z\"{u}rich, Switzerland}

\author{J.~K\"{o}hler}
\affiliation{Max-Planck-Institut f\"{u}r Festk\"{o}rperforschung, Heisenbergstr. 1, D-70569 Stuttgart, Germany}

\author{A.~Bussmann-Holder}
\affiliation{Max-Planck-Institut f\"{u}r Festk\"{o}rperforschung, Heisenbergstr. 1, D-70569 Stuttgart, Germany}

\begin{abstract}

 EuTiO$_{3}$ undergoes a structural phase transition from cubic to tetragonal at $T_{\rm S}$ = 282 K which 
is not accompanied by any long range magnetic order. However, it is related to the 
oxygen ocathedra rotation driven by a zone boundary acoustic mode softening. 
Here we show that this displacive second order structural phase transition can 
be shifted to higher temperatures by the application of an external 
magnetic field (${\Delta}$$T_{\rm S}$ ${\simeq}$ 4 K for $\mu_{0}$$H$ = 9 T). 
This observed field dependence is in agreement with theoretical 
predictions based on a coupled spin-anharmonic-phonon interaction model.

\end{abstract}


\maketitle

In the search for novel multiferroic materials the work of Katsufuji and Takagi \cite{Katsufuji} on EuTiO$_{3}$ 
(ETO) has recently invoked hopes in having a new compound with the desired properties. They 
demonstrated that the dielectric constant of ETO increases steadily with decreasing temperature 
to show a dramatic decrease upon the onset of the antiferromagnetic phase transition at $T_{\rm N}$ = 5.5 K. 
This unexpected behavior can be reversed by the application of a magnetic field and clearly 
demonstrates that a strong spin-lattice coupling is present in the system. However, a 
ferroelectric phase transition is absent, since not only quantum fluctuations suppress 
the complete softening of the ferroelectric soft mode \cite{Kamba,Goian}, but also its energy is too 
large to find a finite extrapolated transition temperature as seen in SrTiO$_{3}$ (STO) \cite{Muller}. 
In spite of this negative result, ETO is an interesting candidate to search for novel spin-lattice 
interaction effects, since opposite to other multiferroics, not the transition metal d-states are
occupied and responsible for the low temperature magnetic properties, but the $A$-site ion in $A$$B$O$_{3}$ 
accounts for the antiferromagnetic phase \cite{Katsufuji}. The close analogy between ETO and STO has recently
been shown \cite{Annette} to be complemented by the fact that both compounds undergo a structural phase 
transition from cubic to tetragonal. While the one of STO takes 
place at $T_{\rm S}$ = 105 K \cite{Muller, Muller2,fleury,Kim,unoki,Kirkp,Shirane,Cow}, 
the corresponding one of ETO occurs at much higher temperatures, namely at $T_{\rm S}$ = 282 K, where the 
space group of the structure changes from $Pm$$\overline{3}$$m$ to $I$4/$mcm$ \cite{kohler,Allieta}. This phase transition has been explored 
by various techniques: specific heat measurements, electron paramagnetic resonance (EPR), muon-spin rotation (${\mu}$SR), 
and X-ray scattering \cite{Annette,kohler,Guguchia,Bussmann}. All experimental results clearly 
evidence the structural instability, and it was also detected in the mixed crystals of Sr$_{1-x}$Eu$_{x}$TiO$_{3}$ \cite{Guguchia}.
 
 An especially interesting result was obtained from ${\mu}$SR experiments where a finite relaxation rate $\lambda_{\rm para}$
could be detected in the paramagnetic phase of ETO up to elevated temperatures even exceeding $T_{\rm S}$ \cite{Bussmann}. 
Its temperature dependence follows closely the one of the soft zone boundary mode as well as the one 
of the inverse EPR line width. This finding not only proves that the spin-lattice coupling is strong,
but also demonstrates that correlated fluctuating spins are present which form finite size clusters. 
Theoretically, this observation has been modeled within a spin-lattice coupled model Hamiltonian 
from which a hybrid paramagnon phonon coupled mode has been 
predicted to appear far above $T_{\rm N}$ \cite{Annette,Guguchia,Bussmann}. 
Another consequence of this approach is that a magnetic field has a substantial influence on the 
eigenfrequencies of the system. This is shown in Fig.~1a where the three frequency branches, optic,
acoustic and paramagnon, are shown for $T$ = 300 K, as a function of the spin-lattice coupling ${\varepsilon}$
which is proportional to the magnetic field $H$ \cite{Jacobsen}. With increasing coupling the zone boundary 
transverse acoustic (TA) mode frequency decreases systematically whereas the transverse optic (TO) 
long wave length mode is not affected by the magnetic field. There is, however, a strong 
paramagnon-optic-phonon coupling at finite momentum q ${\simeq}$ 0.25 which suggests the formation of 
lattice mediated spin cluster formation. 
 
 The influence of the magnetic field on the acoustic mode is not only accompanied by 
its softening at the zone boundary, but substantial anomalies are also seen in the 
long wave length limit, pointing to a pronounced softening of the elastic constants 
under an applied magnetic field, namely strong magneto-elastic coupling which is shown 
in Fig.~1b where the acoustic mode dispersion is normalized to its zero field 
dispersion. For small field strengths this coupling is rather inefficient, 
but becomes distinct with increasing field. 
   
%%%%%%%%%%%%%%%%%%%%%%%%%%%%%%%%%%%%%%%%%%%%%%%%%%%%%%%%%%%%%%
\begin{figure}[t!]
\includegraphics[width=1.1\linewidth]{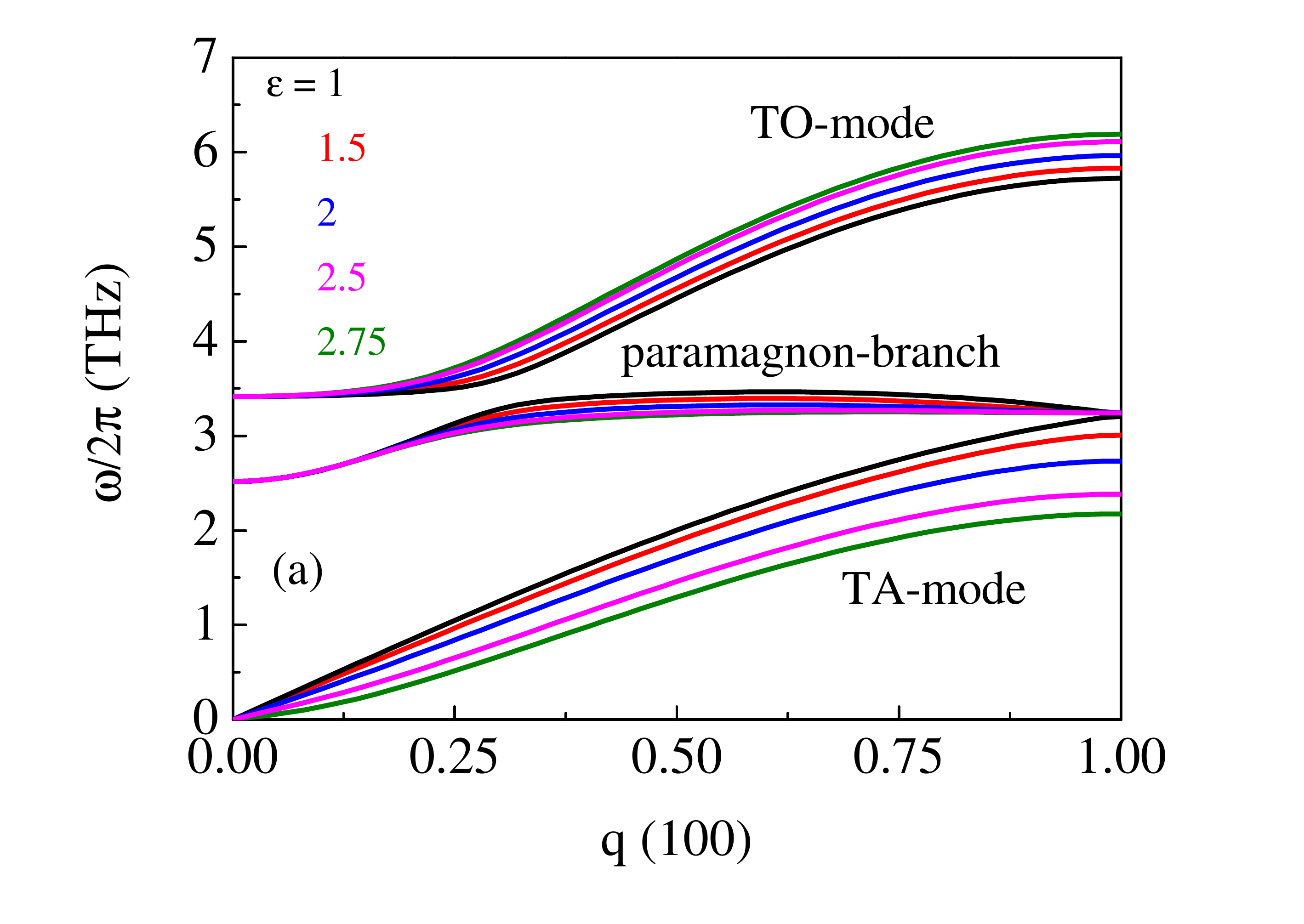}
\includegraphics[width=1.1\linewidth]{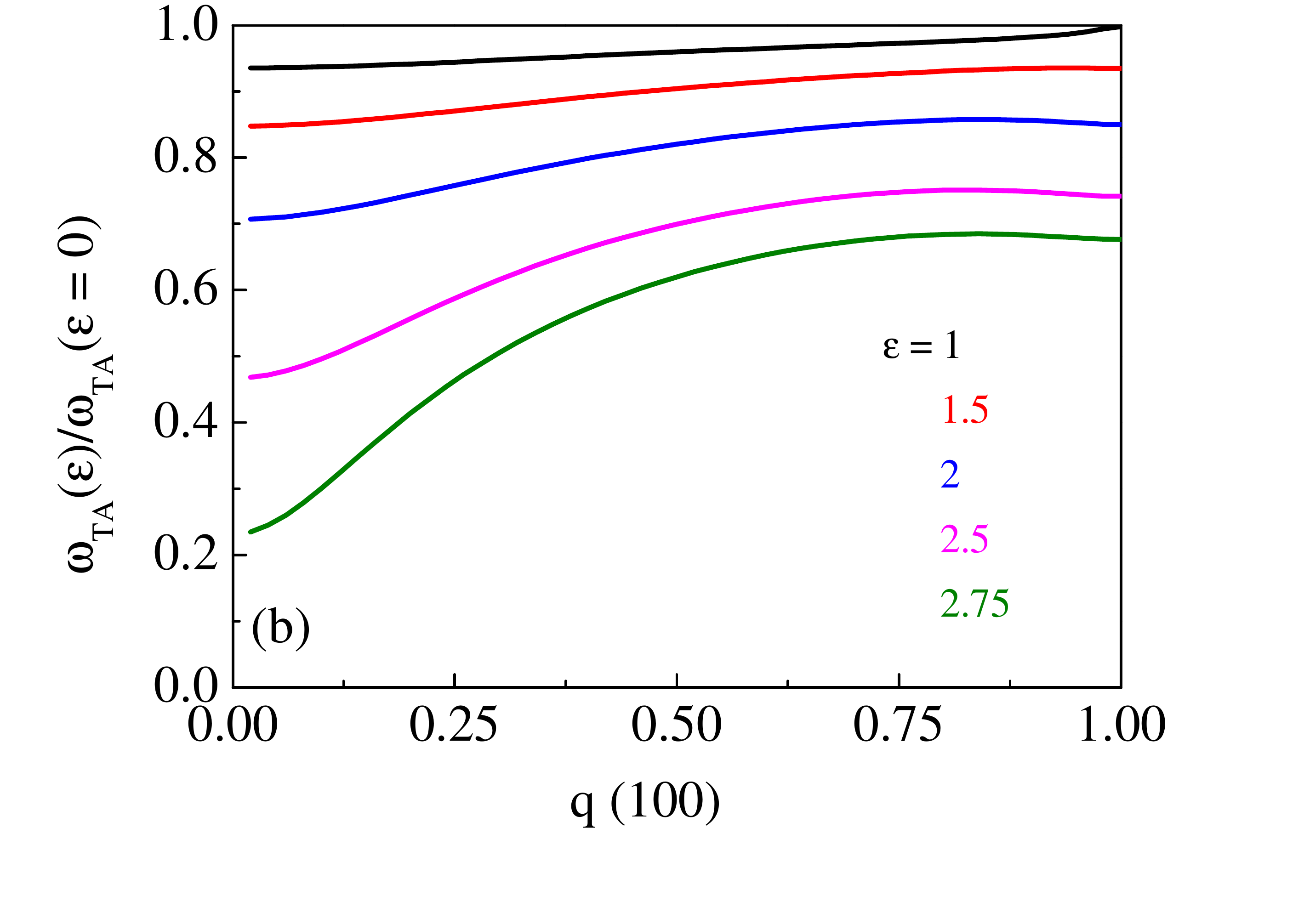}
\vspace{-1.0cm}
\caption{ (Color online) a) Dispersion of the acoustic, optic and paramagnon modes 
at $T$ = 300 K for different coupling strengths ${\varepsilon}$ ${\propto}$ $H$ as indicated in the figure. 
b) Acoustic mode dispersion normalized to the zero field 
dispersion for the same values of ${\varepsilon}$ as in the main part of the figure.}
\label{fig1}
\end{figure}
%%%%%%%%%%%%%%%% 
 The softening of the zone boundary mode frequency with increasing field is 
an indirect proof that the structural phase transition temperature $T_{\rm S}$ is also 
field dependent. This has been tested by performing specific heat measurements 
in an applied magnetic field over a broad temperature regime including the N{\'{e}}el phase. 

 The sample preparation has been described in Ref.~5. The specific heat
measurements (relaxor type calorimeter 
Physical Properties Measurements System $Quantum$ $Design$) 
were first performed in zero field and then repeated in fields of 3, 7, and 9 T. The results are shown in Fig.~2. 
%%%%%%%%%%%%%%%%%%%%%%%%%%%%%%%%%%%%%%%%%%%%%%%%%%%%%%%%%%%%%%
\begin{figure}[t!]
\includegraphics[width=1.01\linewidth]{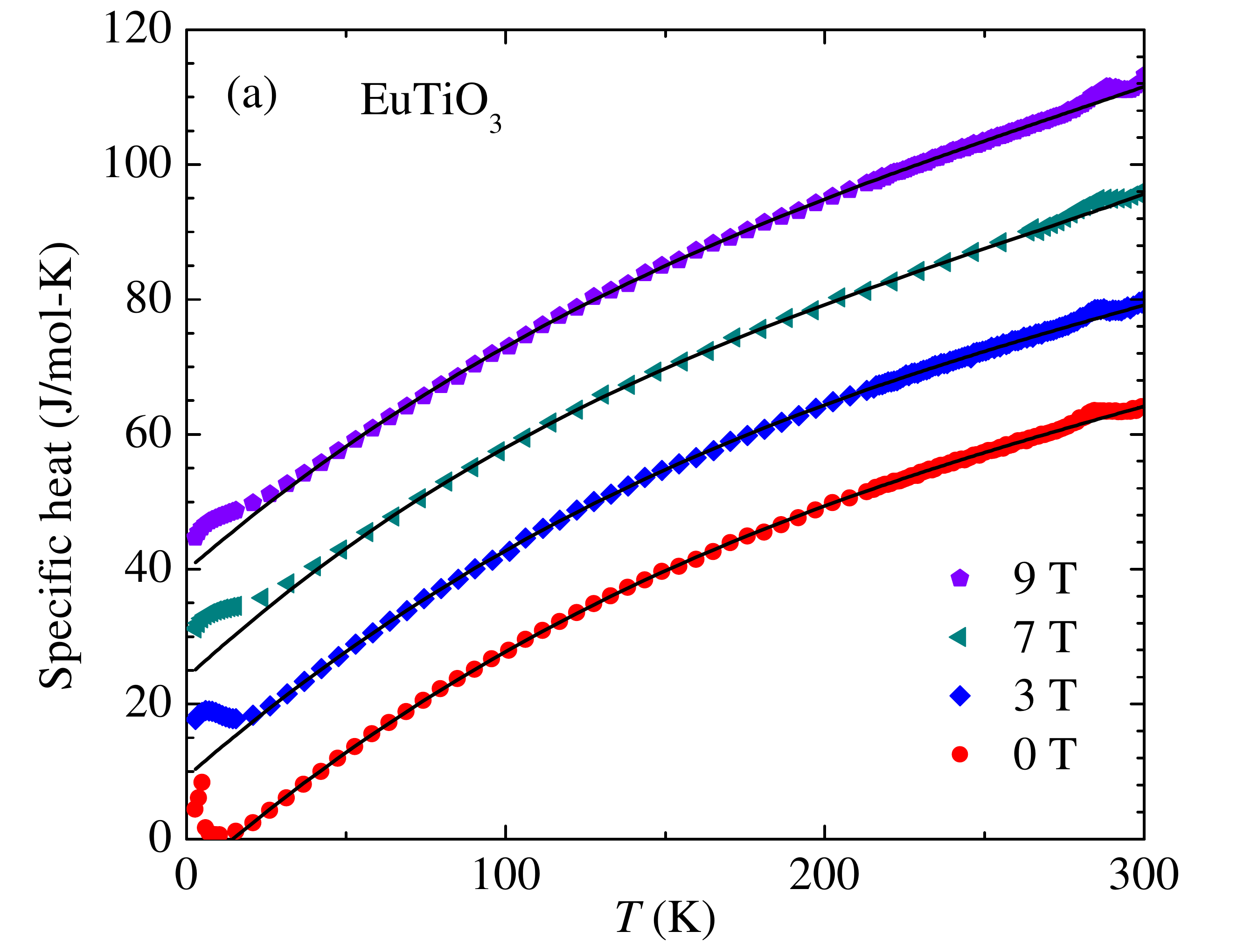}
\includegraphics[width=1.06\linewidth]{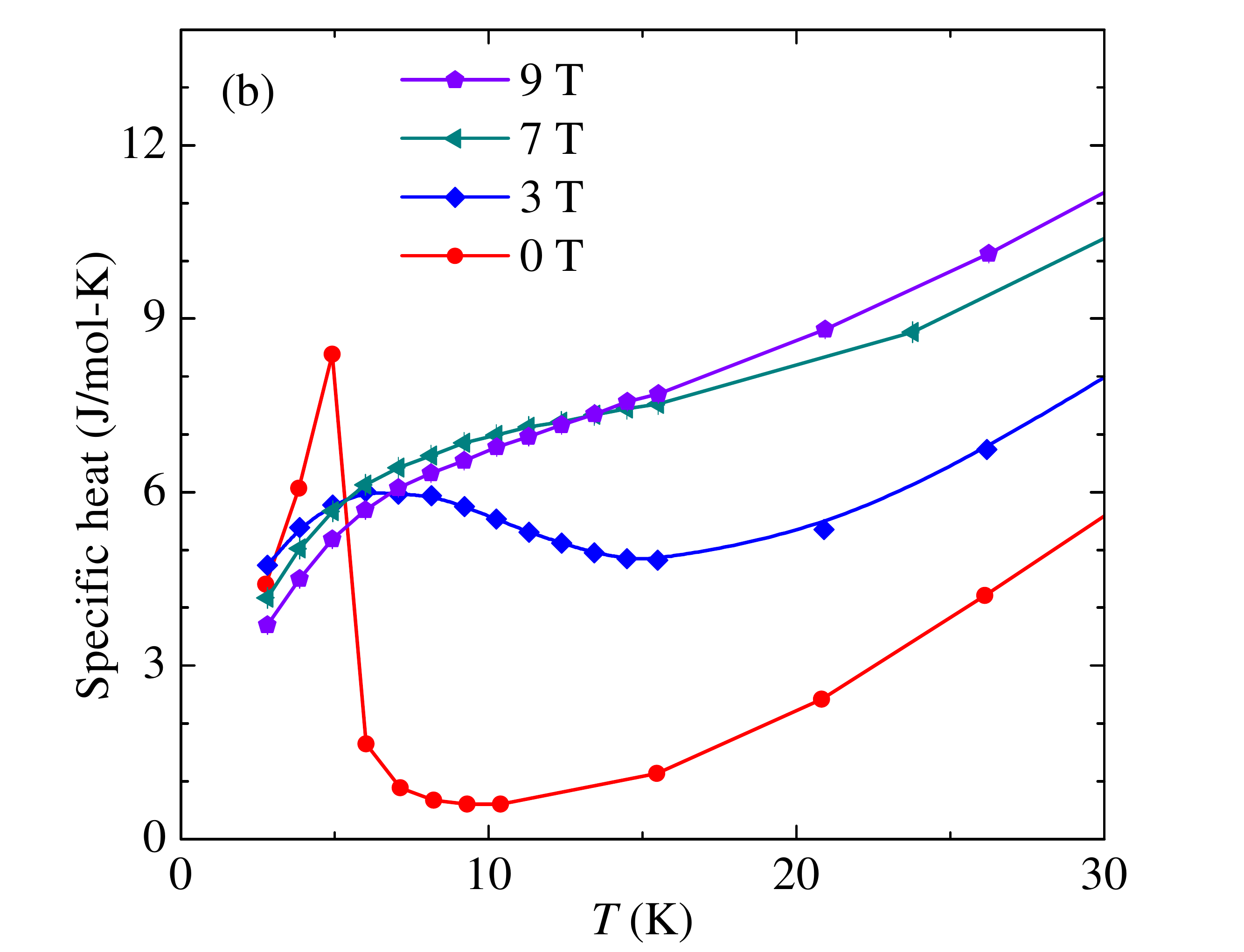}
\vspace{-0.5cm}
\caption{ (Color online) a) Specific heat of ETO as a function of temperature 
at $\mu_{0}$$H$ = 0 T, 3 T, 7 T, and 9 T. For clarity the data are shifted by 15 J/mol-K relative
to each other with increasing magnetic field. b) Specific heat of ETO at low temperatures showing the anomaly at 
$T_{\rm N}$ = 5.5 K (same data as in Fig.~2a).}
\label{fig1}
\end{figure}
%%%%%%%%%%%%%%%% 
%%%%%%%%%%%%%%%%%%%%%%%%%%%%%%%%%%%%%%%%%%%%%%%%%%%%%%%%%%%%%%
\begin{figure}[t!]
\includegraphics[width=1.12\linewidth]{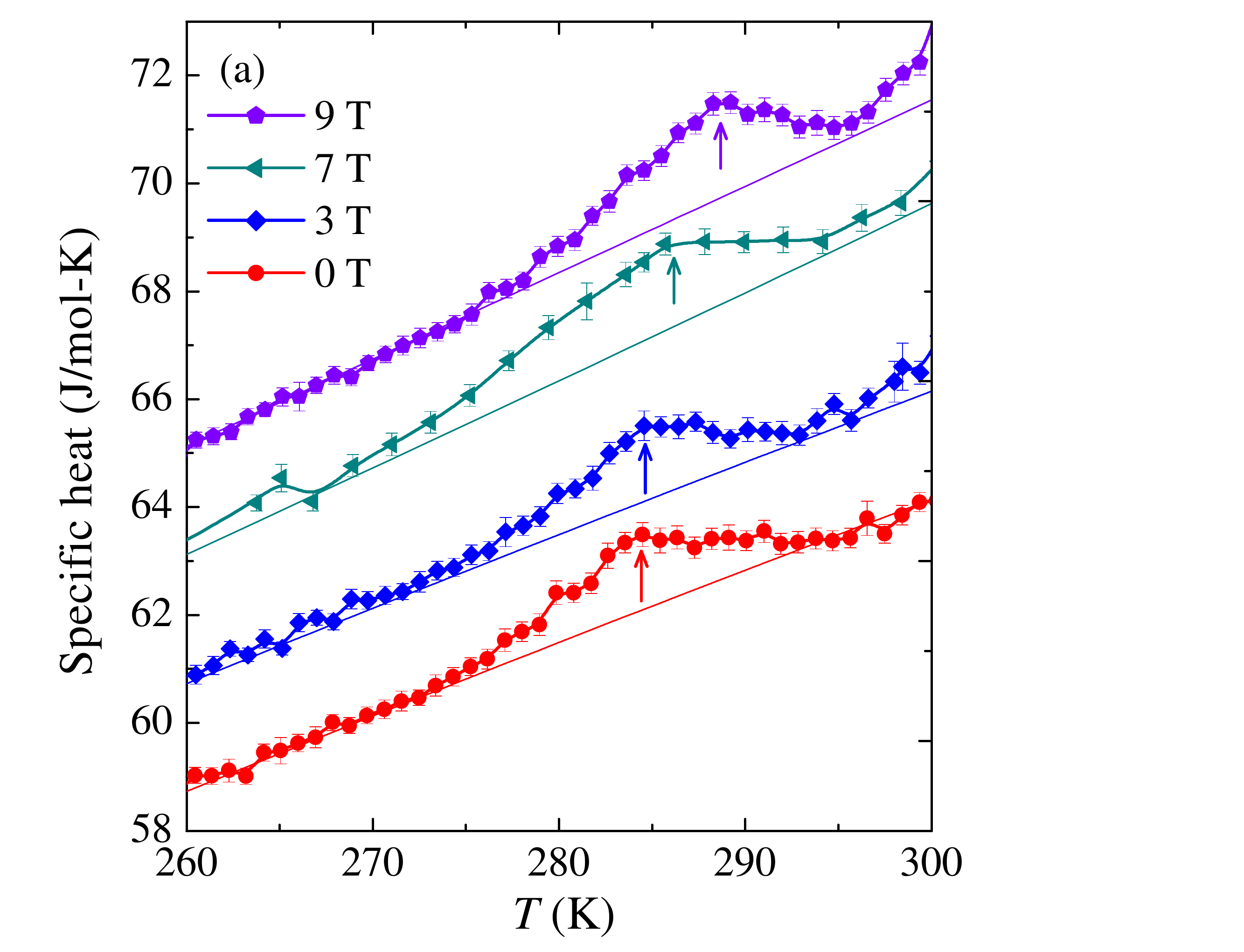}
\includegraphics[width=1.12\linewidth]{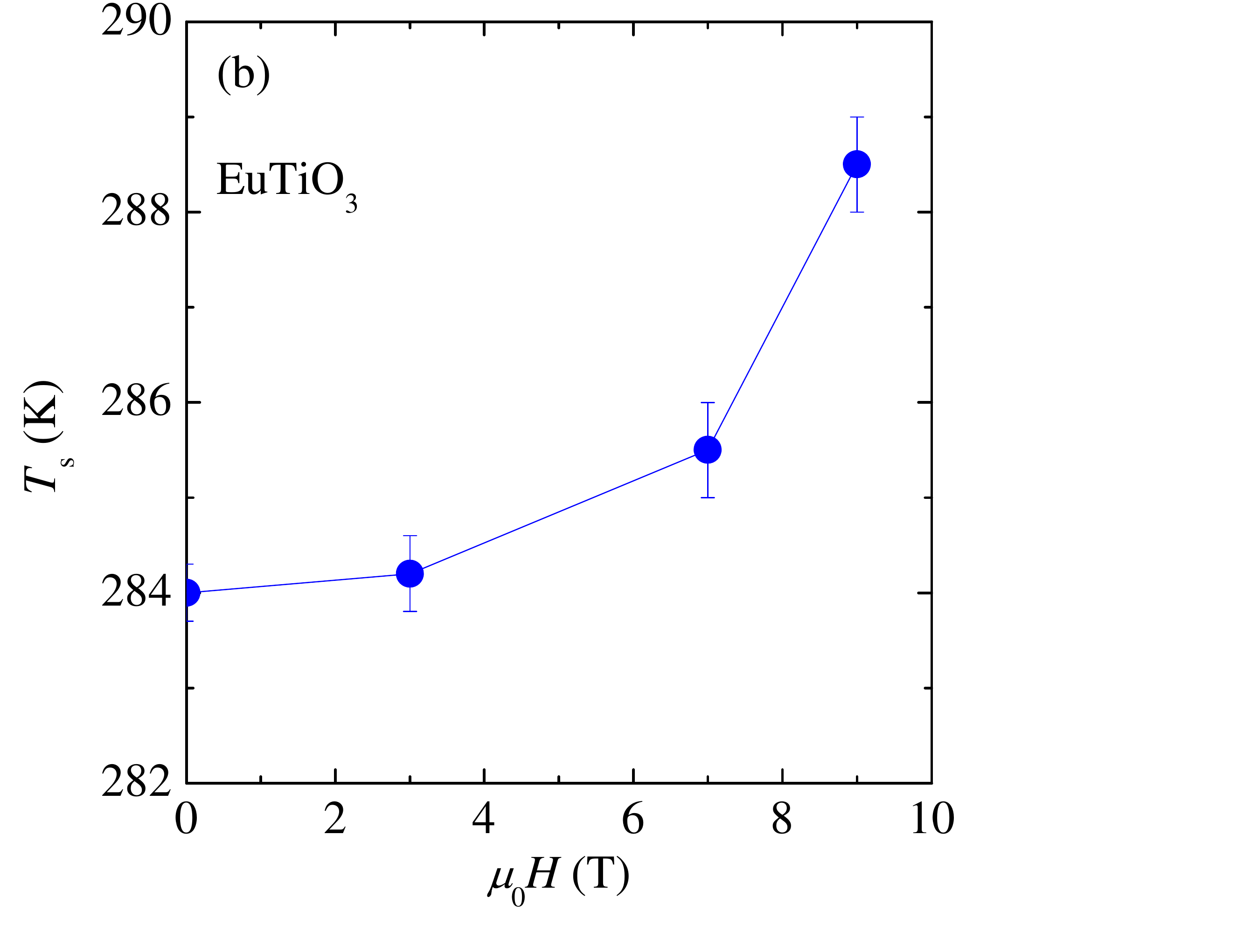}
\vspace{-0.5cm}
\caption{ (Color online) a) The specific heat of ETO within a limited temperature 
regime around the structural phase transition temperature $T_{\rm S}$ for fields of 0, 3, 7, 9 T.
For clarity the data are shifted by 2 J/mol-K relative to each other with increasing magnetic field. 
b) $T_{\rm S}$ as a function of $H$ as derived from Fig.~3a.}
\label{fig1}
\end{figure}
%%%%%%%%%%%%%%%%%%%%%%%%%%%%%%%%%%%%%%%%%%%%%%%%%%%%%%%%%%%%%%
 Importantly, the zero field data (Fig.~2a) are in excellent agreement with our previously 
reported data \cite{Annette}. For the field dependent data it is obvious from Figs.~2a and 2b that the 
low temperature specific heat anomaly caused by the antiferromagnetic phase transition diminishes 
with increasing field in accordance with the dielectric data \cite{Katsufuji} and field dependent specific heat
data \cite{Petrovich}. At high temperatures the anomaly stemming from the structural phase transition is visible as a distinct peak which is in 
the focus of the following discussion (Fig.~2a). By concentrating on a limited temperature range around $T_{\rm S}$ 
the anomaly is enlarged and better visible as shown in Fig.~3a. Obviously a shift of $T_{\rm S}$ to higher temperatures 
takes place with increasing $H$ which is shown in Fig.~3b. While the data of 0 T and 3 T are almost identical,
a nonlinear enhancement of $T_{\rm S}$ is observed for larger magnetic fields. While it is well known, though 
rather rarely observed, that structural phase transitions involving a transition to a magnetically 
ordered phase can be tuned by a magnetic field, the field dependence of the purely structural transition 
of ETO without involvement of magnetic ordering is new. It has also no analogies to other incipient 
ferroelectric perovskites, where the oxygen octahedral rotation instability is insensitive to a magnetic 
field. The consequences are multifold since a tuning of piezoelectric and pyroelectric effects in 
amorphous ETO analogous to STO and BaZrO$_{3}$ \cite{Ehre,Frenkel} is possible by a magnetic field and 
novel functionalities can be expected. 
%%%%%%%%%%%%%%%%%%%%%%%%%%%%%%%%%%%%%%%%%%%%%%%%%%%%%%%%%%%%%%
\begin{figure}[t!]
\includegraphics[width=1.0\linewidth]{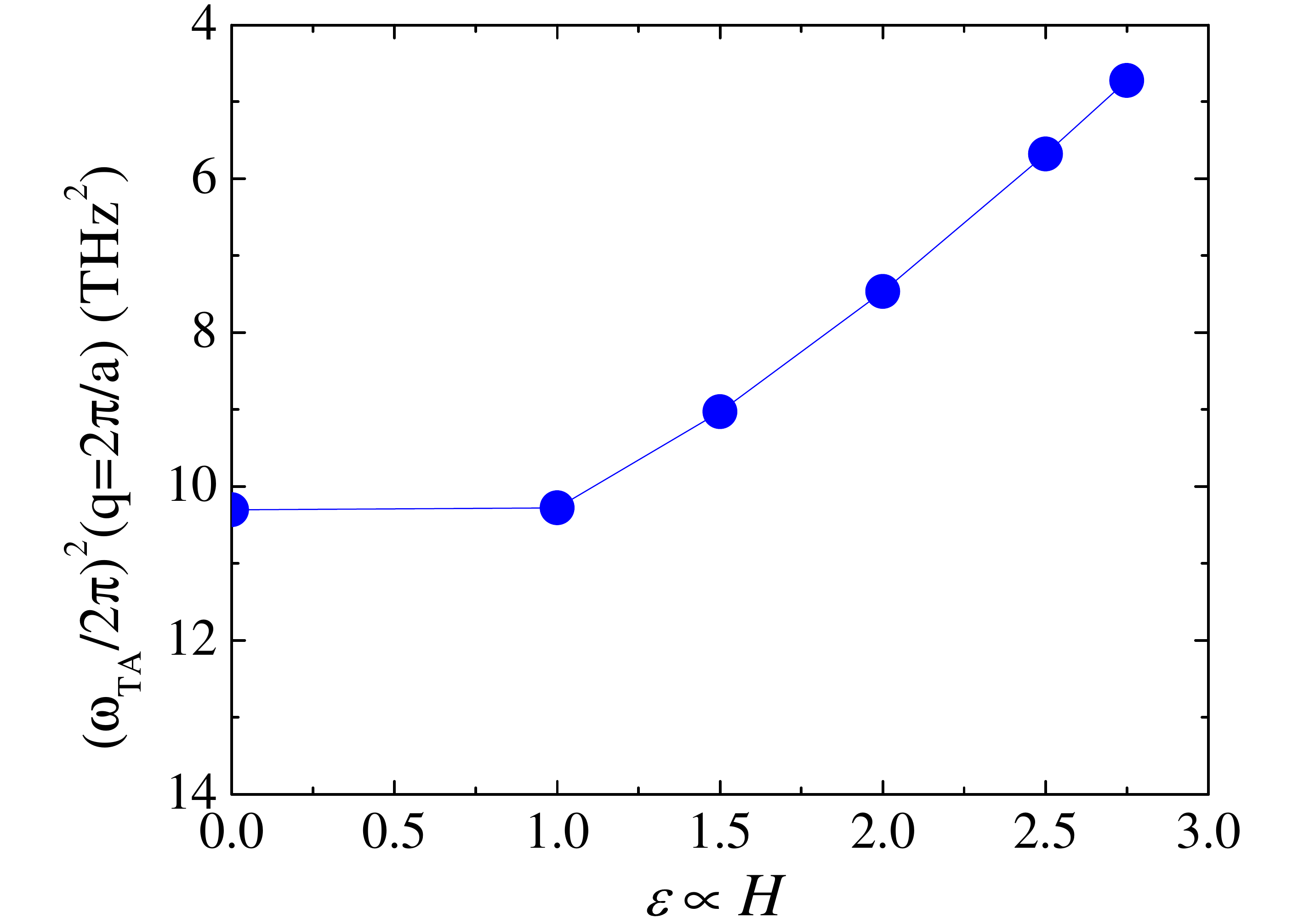}
\vspace{-0.5cm}
\caption{ (Color online) Field dependence of the acoustic zone 
boundary mode frequency ${\omega}^2_{\rm TA}$ ($q$ = 2${\pi}$/$a$) as derived from Fig.~1.}
\label{fig1}
\end{figure}
%%%%%%%%%%%%%%%%%%%%%%%%%%%%%%%%%%%%%%%%%%%%%%%%%%%%%%%%%%%%%%
 The experimental data can be compared to the theoretically 
derived ones by plotting the zone boundary acoustic mode as a function of ${\varepsilon}$ (Fig.~4). 
As is obvious from Fig.~4 a nonlinear dependence of ${\omega}^2_{\rm TA}$ ($q$ = 2${\pi}$/$a$) on ${\varepsilon}$
is obtained with an analogous dependence of the frequency on the 
coupling as $T_{\rm S}$ on $H$ (Fig. 3b). Importantly, the zone boundary 
acoustic mode decreases with increasing field indicating the increased tendency towards the 
structural instability in accordance with the experiment (Figs.~3a and 3b).
The observed magnetic field dependence of the structural instability cannot 
be explained by the fact that the phase transition is decoupled from the Eu 
spins which do not show any long range order in this temperature regime. 
Instead and in accordance with our previous data and analysis \cite{Bussmann}, 
it must be concluded that finite size spin ordered clusters exist in ETO 
at high temperatures which strongly interact with the lattice dynamics. 
Since it is well known for STO that dynamical precursor effects are present 
above the structural phase transition temperature which increase in size 
upon approaching $T_{\rm S}$ \cite{Roleder} we can assume also for ETO that similar dynamics 
are realized here as well. However, these finite size clusters obviously 
couple to the spin subsystem and induce local spin ordering. 
The magnetic field effect contributes to their spatial extent 
and influences thereby the growth of the precursor dynamics. 
As such a novel and interesting interplay between the lattice 
and the spins takes place which to our knowledge has not been reported before. 
Another consequence of our combined experimental - theoretical approach is the 
observation that the magnetic field has pronounced effects on the elastic 
properties of ETO, namely a strong elastic softening. This is in contrast 
to temperature effects where no such softening is present \cite{Bettis}, suggesting 
that ETO should exhibit a huge magneto-elastic coupling effect leading to novel functionalities. 

 In conclusion, we have shown that the structural phase transition of ETO 
can be manipulated by an external magnetic field which shifts $T_{\rm S}$ to higher 
temperatures with the dependence of $T_{\rm S}$ on $H$ being nonlinear. Theoretically 
a similar dependence of the zone boundary transverse mode frequency on the 
spin-lattice coupling is found where an increased softening with increasing 
coupling strength ${\varepsilon}$ takes place. The additional implications are large magneto-elastic 
interactions which should be observable by resonant ultrasound spectroscopy. 
The experimental data together with their interpretation suggest that dynamical 
precursor clusters form above the actual structural phase transition temperature 
which are tied to the Eu spins to induce a locally ordered spin alignment via the 
bridging oxygen ions thus possibly leading to ferromagnetic domains through the superexchange mechanism [21].

%%%%%%%%%%%%%%%%%
\section{Acknowledgments}~
~This work was supported by the Swiss National Science Foundation and the
SCOPES grant No.IZ73Z0${\_}$128242.

\end{document}